\documentclass[fleqn,10pt]{wlscirep}
\usepackage[utf8]{inputenc}
\usepackage[T1]{fontenc}
\usepackage{multirow}

\title{A tissue and cell-level annotated H\&E and PD-L1 histopathology image dataset in non-small cell lung cancer}

\author[1,$\dag$]{Joey Spronck}
\author[1,$\dag$]{Leander van Eekelen}
\author[1]{Dominique van Midden}
\author[1]{Joep Bogaerts}
\author[1]{Leslie Tessier}
\author[1]{Valerie Dechering}
\author[1]{Muradije Demirel-Andishmand}
\author[1]{Gabriel Silva de Souza}
\author[1]{Roland Nemeth}
\author[2]{Enrico Munari}
\author[3]{Giuseppe Bogina}
\author[4]{Ilaria Girolami}
\author[5]{Albino Eccher}
\author[6]{Balazs Acs}
\author[6]{Ceren Boyaci}
\author[7]{Natalie Klubickova}
\author[1]{Monika Looijen-Salamon}
\author[1]{Shoko Vos}
\author[1,*]{Francesco Ciompi}
\affil[1]{Department of Pathology, Radboud University Medical Center, Nijmegen, The Netherlands}
\affil[2]{Pathology Unit, Department of Molecular and Translational Medicine, University of Brescia, Brescia, Italy}
\affil[3]{Department of Pathology, Ospedale Sacro Cuore, Negrar, Verona, Italy}
\affil[4]{Department of Pathology, Provincial Hospital of Bolzano (SABES-ASDAA), Bolzano-Bozen, Italy}
\affil[5]{Department of Pathology and Diagnostics, University and Hospital Trust of Verona, Verona, Italy}
\affil[6]{Department of Clinical Pathology and Cancer Diagnostics, Karolinska University Hospital, Stockholm, Sweden}
\affil[7]{Biopticka Laboratory, Ltd, Pilsen, Czech Republic}
\affil[*]{Corresponding author(s): Joey Spronck (joey.spronck@radboudumc.nl), Leander van Eekelen (leander.vaneekelen@radboudumc.nl), Francesco Ciompi (francesco.ciompi@radboudumc.nl)}
\affil[$\dag$]{These authors contributed equally to this work.}

\begin{abstract}
The tumor immune microenvironment (TIME) in non-small cell lung cancer (NSCLC) histopathology contains morphological and molecular characteristics predictive of immunotherapy response. Computational quantification of TIME characteristics, such as cell detection and tissue segmentation, can support biomarker development. However, currently available digital pathology datasets of NSCLC for the development of cell detection or tissue segmentation algorithms are limited in scope, lack annotations of clinically prevalent metastatic sites, and forgo molecular information such as PD-L1 immunohistochemistry (IHC). To fill this gap, we introduce the \textit{IGNITE data toolkit}, a multi-stain, multi-centric, and multi-scanner dataset of annotated NSCLC whole-slide images. We publicly release 887 fully annotated regions of interest from 155 unique patients across three complementary tasks: (i) multi-class semantic segmentation of tissue compartments in H\&E-stained slides, with 16 classes spanning primary and metastatic NSCLC, (ii) nuclei detection, and (iii) PD-L1 positive tumor cell detection in PD-L1 IHC slides. To the best of our knowledge, this is the first public NSCLC dataset with manual annotations of H\&E in metastatic sites and PD-L1 IHC.

\end{abstract}
\begin{document}

\flushbottom
\maketitle
\thispagestyle{empty}

\section*{Background \& Summary}

Immune checkpoint inhibitors (ICIs) have made a significant impact on the treatment of both early and late-stage non-small cell lung cancer (NSCLC) patients. In late-stage NSCLC, first-line (combination) ICI therapy has started to become common practice \cite{reck2022first} with significant survival benefits over chemotherapies \cite{reck2021five, sezer2021cemiplimab, herbst2021fp13}, while in early stage NSCLC, (neo)adjuvant ICI treatment \cite{meng2024efficacy} is gradually being integrated into American and European clinical guidelines. Despite these advances across disease stages, the overall response rate among ICI-treated patients remains low (approximately 40\% \cite{reck2022first}), indicating that only a subset of patients derive clinical benefit from them. This is partly due to the poor predictive power of the biomarker currently used in the clinic for treatment selection \cite{yang2021comparative}, the tumor proportion score (TPS), based on PD-L1 immunohistochemistry (IHC) as the fraction of PD-L1-positive tumor cells over all tumor cells. Therefore, there exists an urgent need for biomarkers that are more predictive of treatment response to ICIs than TPS, to increase the proportion of patients who benefit from treatment, to save patients from treatment-related adverse effects, and to improve the overall cost-effectiveness of ICIs.

Histopathology serves a crucial role in the clinical assessment of NSCLC. At a basic level, hematoxylin \& eosin (H\&E) stained slides offer insight into tissue morphology and structure, while PD-L1 IHC provides information on the immune checkpoint mechanism, the effect of which is currently summarized into the TPS. Furthermore, histopathology offers the opportunity to study the tumor immune microenvironment (TIME) \cite{binnewies2018understanding}, the interaction between the tumor, the immune system, and the surrounding tissues. Aspects of the TIME such as the tumor-infiltrating lymphocytes (TILs), necrosis or tertiary lymphoid structures (TLS) have been shown to be predictive for immunotherapy outcome \cite{wang2022characteristics, tilguidelines2}, but their visual quantification can be difficult due to the associated inter-rater variability \cite{kos2020pitfalls, sato2023roles, tilguidelines1, tilguidelines2}, suggesting these biomarkers could benefit from quantification with advanced image analysis powered by artificial intelligence (AI).

With the introduction of digital pathology, AI methods such as deep learning can now be developed to analyze whole-slide images (WSIs) \cite{niazi2019digital} and quantify biomarkers such as TILs \cite{backman2023spatial, park2022artificial, spronck2022esmo}, tumor necrosis \cite{kludt2024next} or TLS \cite{van2024multi}. These models are often based on cell detection or tissue segmentation as a first step, thereby counting, quantifying, and analyzing the morphology and spatial interaction of cells. These detections and segmentations can later serve as input for downstream classification models. 

Segmentation and detection models are usually trained in a fully supervised manner, requiring large amounts of manually annotated data.
However, publicly available NSCLC histopathology datasets for tasks such as nuclei detection or tumor versus benign tissue segmentation are scarce and limited in scope \cite{kludt2024next, li2020deep, verma2021monusac2020, rkaczkowska2022deep, han2022wsss4luad}.  While large and generic histopathology datasets for tissue segmentation and cell detection exist (such as SegPath \cite{komura2023restaining}), previous studies such at the MIDOG challenge \cite{aubreville2023mitosis} show that the domain shift of moving to an organ outside of the training set leads to significant performance loss, warranting organ-specific datasets to train on. Moreover, the pre-existing NSCLC datasets are limited to H\&E slides; to the best of our knowledge, no annotated datasets exist for cell-level PD-L1 annotations, limiting the potential of deep learning models for biomarker development on IHC slides. Lastly, approximately 40 percent of NSCLC patients present with distant metastasis at diagnosis, most frequently to the liver, bone, brain, and adrenal glands \cite{riihimaki2014metastatic, tamura2015specific}. However, metastatic NSCLC cases are completely absent in these datasets, limiting the applicability of the developed models to these clinically prevalent metastatic cases.

To address the gaps in existing data, we introduce the \emph{IGNITE data toolkit}, a multi-stain, multi-centric, and multi-scanner dataset of annotated digital pathology images for the analysis of H\&E- and PD-L1-stained WSIs in NSCLC. This data is the expanded version of unreleased data previously used in Spronck et al. \cite{spronck2023nnunet} and Van Eekelen et al. \cite{van2024comparing}.  The toolkit features three complementary datasets in NSCLC histopathology: i) multi-class semantic segmentation of tissue compartments in H\&E-stained slides, ii) the detection of nuclei in IHC, and iii) the detection of PD-L1-positive tumor cells in PD-L1 IHC slides. The H\&E annotations can power a detailed analysis of NSCLC morphology with 16 classes, including tissue compartments that are relevant for the analysis of the TIME such as tumor cells, stroma, inflamed regions, necrotic regions, and macrophages \cite{Salgado2015til, park2022artificial, sedighzadeh2021macrophages}, in line with the TIL biomarker guidelines developed by the International ImmunoOncology Biomarker Working Group \cite{tilguidelines1, tilguidelines2}. In addition, frequent NSCLC metastatic sites like the liver and the brain were annotated to allow deep learning models to generalize to non-lung morphologies. Importantly, to the best of our knowledge, this is the first public release of cell-level PD-L1 IHC annotations; we release PD-L1 annotations made across three PD-L1 monoclonal antibodies from two different clinical centers.

With the IGNITE data toolkit, we aim to provide a publicly available resource to develop deep learning models to improve tissue segmentation and cell detection in NSCLC, to enable accurate downstream quantification of the TIME, and stimulate the development of novel biomarkers for ICI treatment response.

\section*{Methods}

In this section, we describe the collection, histological preparation, digitization, and annotation process of the cases included in each of the datasets. We also detail the procedure to train deep learning models in a fully supervised manner using each of the introduced datasets, which serves both as a technical validation and as an example of a practical use case of the proposed data toolkit. A global overview of our data collection and annotation methods is depicted in Figure \ref{fig:methods}.

\begin{figure}
    \centering
    \includegraphics[width=0.82  \linewidth]{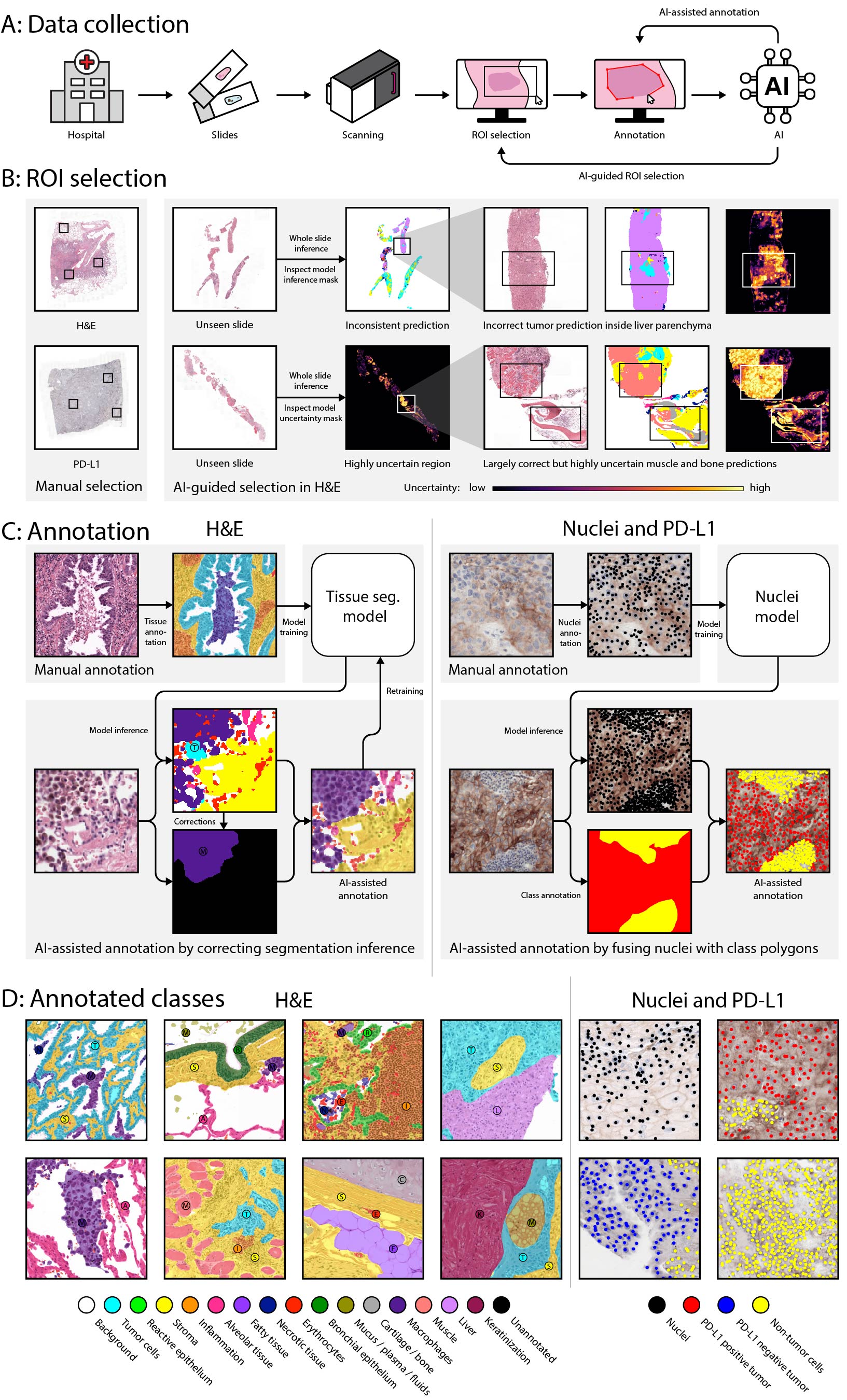}
\end{figure}
\begin{figure}
    \caption{A summary of the data collection and (AI-assisted) annotation process. In \textbf{(A)}, we schematically represent how we use a human-in-the-loop workflow for our annotation, where regions of interest (ROIs) and annotations are selected/corrected based on the performance of intermediate deep learning models. \textbf{(B)} shows the procedures for ROI selection, where `AI-guided' ROI selection leveraged AI output to identify new ROIs for targeted annotation. 
    In \textbf{(C)} we show details on the annotation procedures, where `AI-assisted' annotations are used to offer time savings over from-scratch manual annotations. For the \textbf{H\&E} dataset we train a preliminary model using an initial batch of manual annotations, and then perform inference on new, unseen regions. For AI-assisted annotations, model inference is merged with sparse annotations to correct and finalize new annotations. For the \textbf{Nuclei and PD-L1} dataset, a nuclei detection algorithm was trained using manual annotations of nuclei. The PD-L1 positive tumor cell detection annotations were made by merging nuclei detections with hand-drawn regions of cells belonging to the same class. \textbf{(D)} shows a representative overview of annotated patches from our datasets with all of the corresponding classes.}
    \label{fig:methods}
\end{figure}

\subsection*{Collection of cases and digitization}
We retrospectively collected H\&E and PD-L1-stained cases from two hospitals: the Radboud University Medical Center, Nijmegen, the Netherlands (referred to as \textit{RUMC}) and the Sacro Cuore Don Calabria Hospital, Verona, Italy (referred to as \textit{SCDC}). For a subset of RUMC cases, we collected additional CD68-stained serial sections to guide the annotations of macrophages in H\&E slides. 
The institutional review boards of the centers waived the need for informed consent (RUMC reference 2018-4764; SCDC reference 2014-005118-49). PD-L1 cases from RUMC were stained with the E1L3N monoclonal antibody (Cell Signaling Technology) and 22C3 (Agilent), while cases from SCDC were stained with 22C3 and SP263 (Ventana). We also used H\&E-stained cases from the Lung Adenocarcinoma (LUAD) and Lung Squamous Cell Carcinoma (LUSC) datasets of The Cancer Genome Atlas (TCGA). 

All histological specimens were cut and stained according to the protocols of their respective hospital's pathology laboratories. They were subsequently digitized using different scanners and resolutions. We collected biopsies, resections, and tissue micro-arrays (TMAs) from various histological NSCLC subtypes. We show an overview of how the collected cases were divided among the three datasets in Figure \ref{fig:data_records}, alongside details such as specimen type, scanner type, and histological subtype. After digitization to WSIs, following previous work \cite{litjens20181399,tigerpaper}, we unified the diverse image formats to a standard multi-resolution TIFF format using 80\% quality JPEG compression to balance image quality and file size.

Trained research assistants under the supervision of pathologists made manual annotations of cells and tissue in the form of points and polygons using the open-source ASAP software \cite{asap} for all three datasets. All annotations were confined to regions of interest (ROIs) in the WSIs. In the following sections, we elaborate on further details of the annotation process per dataset. 

\subsection*{Annotations for H\&E tissue segmentation dataset}
We defined 16 classes to categorize several histological regions in the morphological NSCLC landscape: tumor cells, stroma, macrophages, inflammation, alveolar tissue, bronchial epithelium, reactive epithelium, necrotic tissue, keratinization, erythrocytes, fatty tissue, cartilage and bone, mucus/plasma/other fluids, muscle, liver parenchyma, background. Following the TIL biomarker guidelines \cite{tilguidelines1, tilguidelines2}, we put particular emphasis on capturing classes relevant for analyzing TIME components, including tumor cells, stroma, inflammation, necrosis, and macrophages. We show graphical examples of all annotated classes in Figure \ref{fig:methods}D. 

In total, 9 experts were involved in annotating the H\&E tissue segmentation dataset; 5 trained annotators (M.V.D.V, K.W, M.D.A, R.N, J.S) under the supervision of 4 pathologists (D.V.M, J.B, L.T, S.V). On average, 4 ROIs were annotated per slide. Inspired by the heuristic employed in the TIGER challenge \cite{tigerpaper}, we selected ROIs to capture the broad diversity of tissue types within WSIs. This commonly included a tumor region, a non-tumor region (e.g., stroma or alveolar tissue), and a region containing other morphologies (e.g., necrosis, cartilage, or fatty tissue). To reflect the heterogeneous and often disorganized nature of NSCLC tissue, we included both canonical and less canonical regions for each class. For instance, the alveolar tissue class comprised both thin epithelial strands (canonical) and thickened, reactive alveolar structures (less canonical).

\subsubsection*{AI-driven iterative annotation process}
Our annotation process followed a two-step annotation workflow to develop and refine training data for model development. First, trained experts manually annotated a set of ROIs, forming the \emph{initial} dataset used for model development. In the second step, AI models trained on this initial dataset were used for `AI-guided ROI selection' (see Figure \ref{fig:methods}B) and `AI-assisted annotation' (see Figure \ref{fig:methods}C). Once new annotations were made, step 2 was repeated to redirect annotation efforts.

For AI model training, we used the \emph{nnUNet for pathology} framework \cite{spronck2023nnunet}, which automatically configures and trains an ensemble of five UNet models using cross-validation (see the `Validation of the IGNITE data toolkit' section for details). As shown in previous work \cite{spronck2023nnunet}, this approach also enables pixel-level inference \emph{uncertainty}, which was used in this study to guide the selection of new ROIs for annotation.

The application of trained AI models to unseen WSIs revealed 4 major types of model limitations: 
i) incorrect class predictions,
ii) regions with high uncertainty,
iii) regions where class boundaries appeared inconsistent,
and iv) missing classes.

These model limitations informed expert annotators in selecting new ROIs for refinement. Specifically, in `AI-guided ROI selection', new ROIs were directly selected where inference masks or uncertainty masks showed model limitations. Figure \ref{fig:methods}B illustrates examples of ROI selection based on inference and uncertainty outputs. Experts either manually annotated full ROIs or further leveraged AI output in `AI-assisted annotation', where sparse corrections were merged with the model predictions to speed up the annotation process (Figure \ref{fig:methods}C). Alternatively, new ROIs were selected and annotated fully manually, e.g., for test set ROIs.

Due to the comprehensive set of defined tissue classes, our dataset did not require an explicit `other' class to capture unassignable regions. However, we used an `unannotated' label for regions that were either unintentionally left without a class assignment or intentionally excluded from annotation. This was essential for omitting highly ambiguous or atypical regions that could introduce noise. Later, this label was used to avoid exhaustive manual annotation of already well-predicted structures (e.g., individual erythrocytes), allowing us to focus on more challenging classes such as macrophages and reactive epithelium. 

\subsubsection*{Tissue class descriptions}
Due to its extensive morphological variation, annotating \textit{tumor cells} was a major focus for capturing the diversity of NSCLC tissue. We covered a wide range of NSCLC's morphologies, including its main subtypes (e.g. adenocarcinoma (AD) and squamous cell carcinoma (SC), and large cell (LC), see Figure \ref{fig:data_records}B) and a diverse set of growth patterns (for AD: lepidic, acinar, papillary, micropapillary, solid, mucinous and non-mucinous; for SC: keratinizing and non-keratinizing). The \textit{stroma} class included tumor-associated stroma, healthy lung stroma, fibrotic stroma, and non-lung stroma, because of visual similarities, inherent ambiguity in defining their boundaries, and context dependence. While stroma may contain a variety of immune cells, stromal regions containing dense clusters of lymphocytes were specifically labeled as \textit{inflammation}. \textit{Macrophages}, a morphologically diverse immune cell type that is often difficult to distinguish in H\&E
, were primarily annotated with guidance from CD68 IHC-stained serial sections. In this process, CD68 positivity on consecutive slides served as a reference for identifying macrophages in the corresponding H\&E images. \textit{Necrotic tissue} is morphologically diverse, and includes dead and dying cells and debris. Individual macrophages within necrotic areas were sometimes included in the necrosis annotations due to their immune context.

Healthy lung tissue included \textit{alveolar tissue} (lung parenchyma) and \textit{bronchial epithelium}. \textit{Reactive epithelium} was defined as a separate class to capture morphological alterations that may resemble tumor cells but lack definitive malignant features. However, it was occasionally challenging to distinguish reactive epithelium from malignant epithelium, particularly in regions where there is a gradient from one tissue type to the other. The \textit{erythrocytes} class captures individual or clusters of red blood cells and larger regions of hemorrhage, while \textit{Mucus/plasma/other fluids} form a small rest class for fluids. The \textit{Fatty tissue} class captures both individual and clusters of adipocytes. In squamous cell NSCLC, \textit{Keratinization}, which may closely resemble necrosis, was annotated as a separate class due to its distinct biological characteristics. 

Biopsies from the metastatic sites, liver, bone, brain, and adrenal glands were included, but the proportion of parenchymal tissue compartments varied across these sites. Brain and adrenal gland biopsies primarily contained tumor and stroma surrounded by widespread necrosis, with little to no non-stromal tissue observed. In contrast, \textit{Liver} parenchyma was present abundantly in liver biopsies and was annotated as a dedicated class, to distinguish it from morphologically similar macrophages and tumor cells. With the addition of bone/meninges metastasis, \textit{bone} was added to \textit{cartilage} into a single class due to the morphological and contextual similarities. Additionally, the inclusion of metastatic biopsies revealed \textit{muscle} in some cases, which was annotated as a separate class. 

\subsection*{Annotations for PD-L1 IHC nuclei \& positive tumor cell detection datasets}
The PD-L1 IHC nuclei detection dataset was annotated by five (D.S, H.Q, S.R, J.K, L.M) trained annotators, and the PD-L1+ tumor cell detection dataset was annotated by three trained annotators (L.M, M.D.A, G.S). Both groups were supervised by two pathologists (E.M, M.L.S). For the selection of ROIs in both datasets, we attempted to select an equal proportion of ROIs in each of the clinically relevant bins of <1\%, 1-49\% and >50\% PD-L1+ tumor range (as visually determined by E.M).

\begin{figure}
    \centering
    \includegraphics[width=1\linewidth]{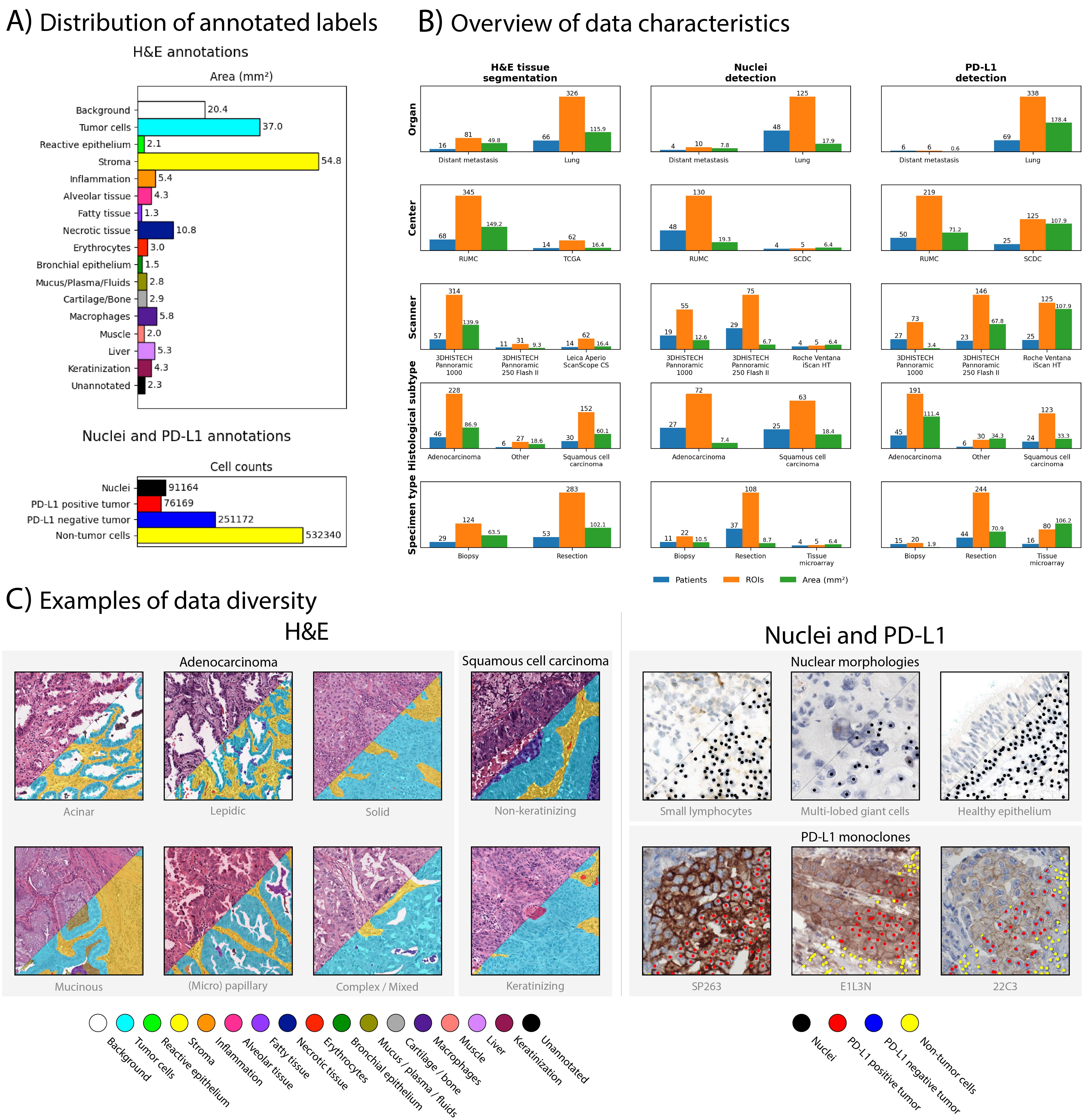}
    \caption{Overview of the three datasets in the IGNITE data toolkit. In \textbf{(A)}, we report the total amount of annotations per class, expressed in square millimeters for the H\&E tissue segmentation dataset and in number of annotated cells for the nuclei/PD-L1+ tumor cell detection datasets. Some regions of interest (ROIs) in the nuclei/PD-L1+ tumor cell detection dataset were annotated by multiple readers; in such cases, we add the mean number of annotations per class to the reported total.  In \textbf{(B)}, we show a quantitative overview of the number of patients, ROIs, and total annotated area per combination of dataset and aspect (organ, source institute, scanner, etc). In \textbf{(C)}, we show graphical examples of the diversity of our dataset, such as differently annotated tumor growth patterns and staining variability (`H\&E'), and nuclei morphologies and PD-L1 stains originating from different monoclonal antibodies (`Nuclei and PD-L1').}
    \label{fig:data_records}
\end{figure}

For the \emph{nuclei detection dataset}, we annotated the centers of all nuclei within the selected ROIs with point annotations. Nuclei whose presence was ambiguous, for instance, because of their faint appearance due to lying deeper in the tissue, were generally not annotated (see Figure \ref{fig:data_records}C for examples). 

For the \emph{PD-L1+ tumor detection dataset}, we annotated three cell types: PD-L1+ tumor cells, PD-L1- tumor cells, and `non-tumor'. PD-L1 positivity was defined as partial or complete circumferential membranous staining above background level. The `non-tumor' type encompassed anything that was not a tumor cell and allowed algorithms trained on the dataset to generalize to other tissues and be applicable to WSIs beyond the tumor bed region. The annotations were made in a two-step process previously proposed in \cite{van2024comparing}: first, the annotators made polygon annotations in the ROIs to label groups of similar cells as one of the three cell types. Secondly, using the nuclei annotations, we developed a nuclei detector for the detection of nuclei centers in PD-L1 IHC (see the `Validation of the IGNITE data toolkit' section for details). This detector was applied to all ROIs, and the detected nuclei were intersected with the polygon annotations and given the corresponding label. We show a schematic representation of this procedure in Figure \ref{fig:methods}C.

\subsection*{Validation of the IGNITE data toolkit}

\subsubsection*{Training}
We trained deep learning models for each of the three datasets, which serves as technical validation of the proposed data toolkit. 

For the H\&E tissue segmentation dataset, we used the \textit{nnUNet-for-pathology} architecture \cite{spronck2023nnunet, isensee2021nnu}, a self-configuring U-Net architecture capable of finding the best possible set of hyperparameters for a particular training set. We trained an nnUNet-for-pathology model using five-fold cross-validation. The nnUNet-for-pathology model was trained using its default settings, without any label weighting or advanced sampling techniques to address class imbalance, as the purpose of this model was to serve as a baseline demonstration rather than an optimized solution. 

For nuclei and PD-L1+ tumor cell detection datasets, we used a training/validation split to train the \textit{YOLOv5} object detection architecture, a detection model shown to have good off-the-shelf performance with minimal hyperparameter tweaking \cite{glenn_jocher_2022_7347926}. For each dataset, we stratified the cross-validation folds or the training/validation sets in terms of histological NSCLC subtype and presence of annotations per class, keeping similar proportions between the folds/the training and validation sets. We show the splits for all three datasets in Figure \ref{fig:technical_validation}A. 

For every dataset, the best-performing models were identified using the validation folds/sets according to the evaluation metrics (see the `Performance metrics' section), and were subsequently applied to hold-out test sets. All networks were trained using 512$\times$512 pixel patches sampled from annotated regions of interest, extracted at 0.5 micrometer/pixel. For additional details regarding the hyperparameters and network architectures of nnUNet and YOLOv5, we refer to Spronck et al. \cite{spronck2023nnunet} and Van Eekelen et al. \cite{van2024comparing}.

\subsubsection*{Testing}
To evaluate the robustness and generalizability of our H\&E segmentation model, we selected a diverse set of cases for the test split. This included i) data from external sites (TCGA cases) with variable staining appearances, scanners, and morphological tumor characteristics (see Figure \ref{fig:data_records}C), ii) non-lung samples, to evaluate robustness to non-lung morphologies such as non-lung stroma and liver parenchyma, and iii) annotations guided by CD68 IHC, to enhance macrophage segmentation evaluation. Importantly, none of the test set annotations were generated using AI-assistance. 

Next to evaluating performance on individual classes, we also assess the model's applicability for segmenting tissue types specifically tailored to TIL analysis. In line with the proposed TIL analysis guidelines \cite{tilguidelines1, tilguidelines2}, test set predictions of the final nnUNet model are grouped into broader categories to evaluate: \textit{tumor cells}, for quantification of intra-tumoral TILs; \textit{stroma and inflammation} grouped, to enable stromal TIL quantification; \textit{macrophages} and \textit{necrosis}, which should both specifically be excluded for TIL analysis; and \textit{rest}. Evaluation with this additional post-processing step serves as a more practical assessment of the model's ability to segment key TIME components for further analysis.

For the holdout test set of nuclei and PD-L1+ tumor detection datasets, we employ a multi-reader setup where each ROI was annotated by multiple readers, so that we could i) measure the human inter-rater variability on both datasets and ii) compare our baseline models to multiple readers. A pathologist (EM) selected ROIs of approximately 150$\times$150 micrometers (4-5 per case) for both test sets using the same criteria as for the training and validation sets. The test set of the PD-L1 nuclei detection dataset was annotated by four readers: three trained student assistants (L.M, M.D.A, and D.Z) and one trained physician (G.S), labeled R\textsubscript{i} through R\textsubscript{iv}. The test set of the PD-L1+ tumor cell detection test set was annotated by two groups, split across hospitals: three expert pathologists (A.E, E.M, and I.G, labeled P\textsubscript{1} through P\textsubscript{3}) read cases from RUMC, and two expert pathologists and one physician (B.A, C.B, and G.S, labeled P\textsubscript{4} through P\textsubscript{6}) read cases from SCDC. Note that model output was not inspected for generating the annotations nor selecting the ROIs of any of the test sets, thus preventing the introduction of selection bias.

To better approximate whole-slide inference conditions and facilitate sliding window approaches, we included additional spatial context surrounding each ROI in all test sets (H\&E, nuclei, and PD-L1 models). This added context ensures that model predictions leverage surrounding tissue architecture, as would occur during full-slide inference. 

We publicly release the model weights, and inference and evaluation pipelines (see the `Code Availability' section). 

\subsubsection*{Performance metrics}
\label{ss:metrics}
We use the F\textsubscript{1} score as an evaluation metric for all datasets. For the H\&E tissue segmentation dataset, the F1 score is calculated per class in a one-versus-rest pixel-wise manner. For the nuclei and PD-L1+ tumor detection datasets, we define a prediction to `hit' a reference standard point if they fall within an $x$ micrometer radius together. For all classes in the PD-L1 positive tumor cell deteciton dataset, we set this radius to 10 micrometer, roughly corresponding to the average diameter of tumor and immune cell nuclei in the dataset. For nuclei in the PD-L1 IHC nuclei detection dataset, we set the stricter radius of 4 micrometers. We define true positives as predictions that hit a reference standard point with the same class label; false positives are predictions that do not hit any reference standard points or are of a different class than the reference standard point; false negatives are missed reference standard points. The standard formula for F1 score can then be applied per class: 2TP/(2TP + FP + FN).

\section*{Data Records}

We release our datasets as a repository on Zenodo \cite{ignitetoolkit}. We provide PNG images of the annotated ROIs cropped to their width and height, extracted at a resolution of 0.5 micrometers per pixel.  We share the annotations for the H\&E tissue compartment segmentation as masks in the form of single-channel PNGs with the same dimensions as the ROIs, where every pixel is labeled with a positive integer value belonging to a certain tissue type. The pixel value-class mapping is available in the \textit{he\_label\_map.json} file. Annotations for the PD-L1 IHC nuclei and positive tumor cell detection datasets are released in the MS COCO format \cite{lin2014microsoft} (see \emph{Usage Notes} section). Lastly, we share a \textit{data\_overview.csv} file of metadata per ROI, e.g. the train/validation/test split in addition to details such as the used scanner and institute the image originated from.

Figure \ref{fig:data_records} provides a quantitative and qualitative overview of the IGNITE data toolkit. In total, 155 unique patients were annotated across all three datasets. Figure \ref{fig:data_records}A shows the distribution of annotated classes in the toolkit, and Figure \ref{fig:data_records}B describes the diversity of the annotated data in terms of number of patients, number of ROIs, and annotated area. For the H\&E tissue segmentation dataset, we annotated 166 mm\textsuperscript{2} of tissue spread over 407 ROIs, focusing foremost on classes that describe aspects of the TIME, such as tumor (22\% of the annotated area), stroma (33\%), necrosis (6\%), and macrophages (3\%). A range of other classes were also annotated to capture the wide variety of tissue morphologies typically encountered in NSCLC cases. Moreover, we also annotated the diverse range of histological growth patterns (see Figure \ref{fig:data_records}C for graphical examples). For the PD-L1 IHC nuclei detection dataset, 91,164 nuclei were annotated over 135 ROIs. We strove to annotate a large variety of nuclei morphologies, ranging from giant multi-lobed examples to elongated fibroblast nuclei or small pneumocyte nuclei (Figure \ref{fig:data_records}C). For the PD-L1+ tumor cell detection dataset, we annotated 859,681 cells over 344 ROIs. While PD-L1+ tumor cells are the smallest class in terms of percentage (8.8\%), this is mostly due to our methodological decision to try to select ROIs `within context', i.e., tumor cells surrounded by stromal components. Nonetheless, the PD-L1 detection dataset shows great variability, for example, in various degrees of cytoplasmic and membranous positivity over the three PD-L1 monoclonal antibodies (Figure \ref{fig:data_records}C).

\section*{Technical Validation}
\begin{figure}
    \centering
    \includegraphics[width=0.95\linewidth]{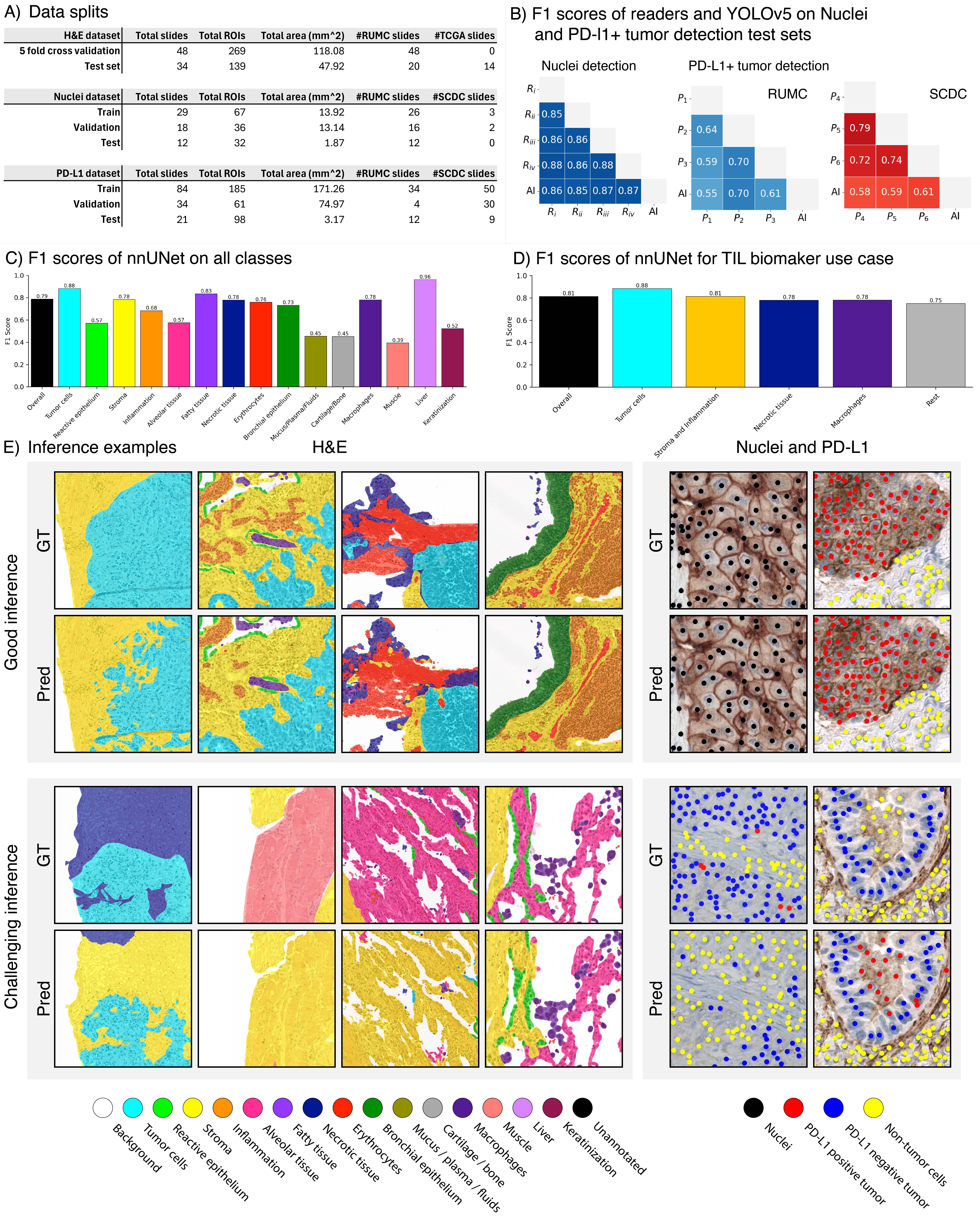}
    \caption{An overview of the technical validation of our datasets. For each dataset, we trained fully-supervised models on top of the data and then evaluated their performance on hold-out test sets. In \textbf{(A)}, we show statistics regarding the train/validation/test split for the baseline models trained on each of the three datasets. In \textbf{(B)}, we show the pairwise F1 scores of readers and the predictions of the algorithms for both the nuclei and PD-L1+ tumor cell detection datasets in PD-L1 IHC. For the PD-L1+ tumor cell detection dataset, we show the F1 scores as averaged over the three classes in the dataset (PD-L1 negative/positive tumor cells and non-tumor cells) and split per clinical center. In \textbf{(C)}, we show the F1 scores of nnUNet on the holdout test set for H\&E. In \textbf{(D)}, we show the same F1 scores, but now grouped according to the biomarker use case. In \textbf{(E)}, we show a quantitative overview of inference from each of the three baseline models.}
    \label{fig:technical_validation}
\end{figure}

In this section, we summarize the performance of our deep learning models on their respective hold-out test sets. We show the data splits and F1 scores, and examples of model inference on the test set for each dataset in Figure \ref{fig:technical_validation}. This performance is meant to show potential use cases and demonstrate the technical validity of the IGNITE data toolkit; we expressly note that our goal in fitting these models was not to reach the highest performance possible, but rather to provide a concrete example of how the annotated data provided in the toolkit can support the training of deep learning models that can be potentially used as building blocks for biomarker development.

\subsection*{H\&E dataset evaluation}
To assess the quality of the H\&E tissue segmentation dataset, performance was evaluated both across all classes and within a subset of tissue classes relevant for TIL quantification, as recommended by current guidelines \cite{tilguidelines1, tilguidelines2} (see Figure \ref{fig:technical_validation}C).

When evaluating all annotated tissue classes, the model achieved an overall F1 score of 0.79. Class-wise performance was highest for liver, tumor cells, fatty tissue, necrotic tissue, stroma, macrophages, and erythrocytes. Moderate F1 scores were observed for bronchial epithelium, inflammation, while lower performance was seen for more challenging or less abundantly annotated classes such as reactive epithelium, alveolar tissue, keratinization, cartilage/bone, mucus/plasma/fluids, and muscle.

For the TIL biomarker use case, classes were merged to reflect the requirements of TIL quantification, such as combining stroma and inflammation into a single category, and grouping all non-relevant classes. In this setup, the model achieved an overall F1 score of 0.81. Here, stroma and inflammation showed an F1 score of 0.81, and the combined `rest' category showed an F1 score of 0.75. This indicates that lower-performing classes in the full evaluation, such as muscle or mucus, do not substantially impact the performance on the use case.

Evaluation of preliminary models highlighted challenges in segmenting reactive epithelium, macrophages, and liver, as these classes were typically misclassified as tumor. Reactive epithelium, even after targeted annotation, remained a challenging morphology and is still largely confused with tumor. This is in line with the diagnostic difficulties in distinguishing these morphologies on H\&E only, especially in the absence of broader tissue context. Similarly, in preliminary models, macrophages were often under-recognized due to their morphological variability in H\&E staining. However, targeted training with and evaluation against IHC-guided annotations yielded an F1 score of 0.78 for macrophages, indicating that the model is capable of learning this class despite its complexity. As is indicated by the F1 score of 0.96, the model learned to identify liver parenchyma, which occurs frequently in metastatic samples.

Examples of segmentation inference are shown in Figure \ref{fig:technical_validation}E. In general, the model performs well on canonical regions of each class. Misclassifications usually occur at boundaries between classes (e.g. where the segmentation model predicted more precise tumor/stroma borders, or differently interpreted the delineation between stroma and inflammation, see inference examples in Figure \ref{fig:technical_validation}E) or regions exhibiting features of multiple tissue types or less canonical regions (e.g., mildly necrotic stroma, or thick compressed alveolar strands in the transition between typical stroma and alveolar tissue, see challenging inference examples in Figure \ref{fig:technical_validation}E). These difficult and less canonical regions hinder overall performance (e.g., the performance on alveolar tissue), while errors in these areas do not always have equal consequences in downstream analysis (e.g., misclassification of stroma as (stromal) inflammation is less critical than misclassifying macrophages as tumor). However, since these less canonical regions are common in NSCLC tissue, we intentionally included this complexity in our dataset. This decision ensures that trained models are exposed to challenging, real-world cases during training and allows for evaluating model behavior under more realistic conditions.

\subsection*{Nuclei and PD-L1 dataset evaluation}
For detecting nuclei in PD-L1 IHC (Figure \ref{fig:technical_validation}B), all pairwise F1 scores between the readers and the algorithms were consistently high (the average reader-reader F1 score and reader-algorithm F1 score were identical at 0.87). This indicates that the algorithm performs within interobserver variability for this test set. The most frequent disagreement between readers was due to nuclei whose presence was ambiguous due to their faint appearance, possibly lying deeper within the tissue slice (see Figure \ref{fig:technical_validation}E for examples). The average number of annotated cells per ROI was 302 $\pm$ 17, while the biggest discrepancy between two readers was 156 cells. For the algorithm and readers, this value was 126. An additional source of disagreement between the algorithm and readers was giant, multi-lobe (cancer) cells (figure \ref{fig:technical_validation}E), where the algorithm predicted each lobe as an independent nucleus.

We show the F1 scores on PD-L1+ tumor cell detection for all reader-algorithm pairings in Figure \ref{fig:technical_validation}B. The F1 scores are shown per clinical center (RUMC, SCDC) and averaged over the three cell types (PD-L1+ tumor cells, PD-L1- tumor cells, and non-tumor cells). Performance on this dataset is generally lower, as indicated by the lower mean reader-reader and reader-algorithm F1 scores (0.70 and 0.61, respectively). When considering the clinical centers separately, the algorithm almost matches the performance of RUMC reader pairs (0.64 versus 0.62) but compares worse against the SCDC reader pairs (0.75 versus 0.59). We show qualitative examples of model inference versus ground truth (GT) in Figure \ref{fig:technical_validation}E. The model frequently considers alveolar macrophages (innately positive for PD-L1) as positive tumor cells, possibly due to a lack of tissue context (seen in the `challenging inference' rows of Figure \ref{fig:technical_validation}E).

\section*{Usage Notes}
For the H\&E tissue segmentation dataset, classes were intentionally split into granular categories. With the TIL biomarker use case as an example, users may choose to group classes according to their specific interests and task requirements. 

For the nuclei and PD-L1 detection test sets, we provide the annotations for all readers to allow comparative studies between AI and experts (see \textit{nuclei\_test\_set\_all\_readers.json} and \textit{'pdl1\_test\_set\_all\_readers.json'}). Moreover, we propose to use a single reader per set as the \emph{canonical} annotator. This canonical annotator functions as a proposed reference standard for future benchmarks and as a way for users of the data to concisely report their own benchmarking results. For this purpose, we choose the three readers who have the best combined ranking of two outcomes: i) highest F1 score among the readers and ii) the highest F1 score versus the respective baseline algorithms. The canonical annotators are R4 for the nuclei detection test set (highest mean reader-reader F1 score: 0.87, tied highest F1 score versus baseline model of 0.87), P2 for the RUMC cases of the PD-L1 detection test set (highest mean reader-reader F1 score: 0.67, highest F1 score versus baseline model: 0.7) and P5 for the SCDC cases (highest mean reader-reader F1 score: 0.765, second highest F1 score versus baseline model: 0.59). We release the training/validation/test set annotations with canonical readers for the nuclei and PD-L1 detection datasets in \textit{'nuclei\_annotations.json'} and \textit{'pdl1\_annotations.json'}).

\section*{Acknowledgements}
We thank Myrthe van de Ven and Kim Wolffenbuttel for helping with the annotation of H\&E cases, and we thank Luca Meesters, Daan Segers, Hiba Qoubbane, Sebastiaan Ram and Joel Käyser for helping with the annotation of PD-L1 IHC cases. This work was supported by a research grant from the NWO (project number 18388).

\section*{Author contributions statement}
J.S, L.V.E and F.C wrote the main manuscript. J.S and L.V.E oversaw the general data collection. D.V.M, J.B, S.V, L.T, V.D, M.D.A and R.N annotated or supervised the annotation of cases for the H\&E tissue segmentation dataset. M.D.A, E.M, I.G, A.E, B.A, C.B, S.V, M.L.S, and N.K annotated or supervised the annotation of cases in the PD-L1 IHC nuclei and positive tumor cell detection datasets. E.M and G.B collected cases from SCDC, and M.L.S and S.V collected cases from RUMC. F.C conceived the study, co-designed experiments, and supervised the work. All authors reviewed the manuscript and approved its final form.

\section*{Competing interests}
I.G is member of the advisory board for Roche Diagnostics. B.A is supported by the Swedish Society for Medical Research postdoctoral grant and by Region Stockholm (clinical research appointment). F.C was chair of the Scientific and Medical Advisory Board of TRIBVN Healthcare, France, and received advisory board fees from TRIBVN Healthcare, France in the last five years. He is shareholder of Aiosyn BV, the Netherlands. All other authors declare no conflict of interest. 

\section*{Code availability}
We provide the code to run test set inference and evaluation on our GitHub repository (\url{https://github.com/DIAGNijmegen/ignite-data-toolkit}). This GitHub repository also contains code to programmatically download all annotated data and trained weights from the Zenodo repository. 

\bibliography{sample}

\begin{thebibliography}{10}
\urlstyle{rm}
\expandafter\ifx\csname url\endcsname\relax
  \def\url#1{\texttt{#1}}\fi
\expandafter\ifx\csname urlprefix\endcsname\relax\def\urlprefix{URL }\fi
\expandafter\ifx\csname doiprefix\endcsname\relax\def\doiprefix{DOI: }\fi
\providecommand{\bibinfo}[2]{#2}
\providecommand{\eprint}[2][]{\url{#2}}

\bibitem{reck2022first}
\bibinfo{author}{Reck, M.}, \bibinfo{author}{Remon, J.} \&
  \bibinfo{author}{Hellmann, M.~D.}
\newblock \bibinfo{journal}{\bibinfo{title}{First-line immunotherapy for
  non--small-cell lung cancer}}.
\newblock {\emph{\JournalTitle{Journal of Clinical Oncology}}}
  \textbf{\bibinfo{volume}{40}}, \bibinfo{pages}{586--597},
  \url{https://doi.org/10.1200/JCO.21.01497} (\bibinfo{year}{2022}).

\bibitem{reck2021five}
\bibinfo{author}{Reck, M.} \emph{et~al.}
\newblock \bibinfo{journal}{\bibinfo{title}{Five-year outcomes with
  pembrolizumab versus chemotherapy for metastatic non--small-cell lung cancer
  with pd-l1 tumor proportion score$\geq$ 50\%}}.
\newblock {\emph{\JournalTitle{Journal of Clinical Oncology}}}
  \textbf{\bibinfo{volume}{39}}, \bibinfo{pages}{2339--2349},
  \url{https://doi.org/10.1200/JCO.21.00174} (\bibinfo{year}{2021}).

\bibitem{sezer2021cemiplimab}
\bibinfo{author}{Sezer, A.} \emph{et~al.}
\newblock \bibinfo{journal}{\bibinfo{title}{Cemiplimab monotherapy for
  first-line treatment of advanced non-small-cell lung cancer with pd-l1 of at
  least 50\%: a multicentre, open-label, global, phase 3, randomised,
  controlled trial}}.
\newblock {\emph{\JournalTitle{The Lancet}}} \textbf{\bibinfo{volume}{397}},
  \bibinfo{pages}{592--604},
  \url{https://doi.org/10.1016/s0140-6736(21)00228-2} (\bibinfo{year}{2021}).

\bibitem{herbst2021fp13}
\bibinfo{author}{Herbst, R.} \emph{et~al.}
\newblock \bibinfo{journal}{\bibinfo{title}{Fp13. 03 impower110: updated os
  analysis of atezolizumab vs platinum-based chemotherapy as first-line
  treatment in pd-l1--selected nsclc}}.
\newblock {\emph{\JournalTitle{Journal of Thoracic Oncology}}}
  \textbf{\bibinfo{volume}{16}}, \bibinfo{pages}{S224--S225},
  \url{https://doi.org/10.1016/j.jtho.2021.01.142} (\bibinfo{year}{2021}).

\bibitem{meng2024efficacy}
\bibinfo{author}{Meng, Y.} \emph{et~al.}
\newblock \bibinfo{journal}{\bibinfo{title}{Efficacy and safety of
  perioperative, neoadjuvant, or adjuvant immunotherapy alone or in combination
  with chemotherapy in early-stage non-small cell lung cancer: a systematic
  review and meta-analysis of randomized clinical trials}}.
\newblock {\emph{\JournalTitle{Therapeutic Advances in Medical Oncology}}}
  \textbf{\bibinfo{volume}{16}}, \bibinfo{pages}{17588359241284929}
  (\bibinfo{year}{2024}).

\bibitem{yang2021comparative}
\bibinfo{author}{Yang, F.}, \bibinfo{author}{Wang, J.~F.},
  \bibinfo{author}{Wang, Y.}, \bibinfo{author}{Liu, B.} \&
  \bibinfo{author}{Molina, J.~R.}
\newblock \bibinfo{journal}{\bibinfo{title}{Comparative analysis of predictive
  biomarkers for pd-1/pd-l1 inhibitors in cancers: developments and
  challenges}}.
\newblock {\emph{\JournalTitle{Cancers}}} \textbf{\bibinfo{volume}{14}},
  \bibinfo{pages}{109}, \url{https://doi.org/10.3390/cancers14010109}
  (\bibinfo{year}{2021}).

\bibitem{binnewies2018understanding}
\bibinfo{author}{Binnewies, M.} \emph{et~al.}
\newblock \bibinfo{journal}{\bibinfo{title}{Understanding the tumor immune
  microenvironment (time) for effective therapy}}.
\newblock {\emph{\JournalTitle{Nature medicine}}}
  \textbf{\bibinfo{volume}{24}}, \bibinfo{pages}{541--550}
  (\bibinfo{year}{2018}).

\bibitem{wang2022characteristics}
\bibinfo{author}{Wang, F.}, \bibinfo{author}{Yang, M.}, \bibinfo{author}{Luo,
  W.} \& \bibinfo{author}{Zhou, Q.}
\newblock \bibinfo{journal}{\bibinfo{title}{Characteristics of tumor
  microenvironment and novel immunotherapeutic strategies for non-small cell
  lung cancer}}.
\newblock {\emph{\JournalTitle{Journal of the National Cancer Center}}}
  \textbf{\bibinfo{volume}{2}}, \bibinfo{pages}{243--262}
  (\bibinfo{year}{2022}).

\bibitem{tilguidelines2}
\bibinfo{author}{Hendry, S.} \emph{et~al.}
\newblock \bibinfo{journal}{\bibinfo{title}{Assessing tumor-infiltrating
  lymphocytes in solid tumors: A practical review for pathologists and proposal
  for a standardized method from the international immuno-oncology biomarkers
  working group: Part 2: {TILs} in melanoma, gastrointestinal tract carcinomas,
  non--small cell lung carcinoma and mesothelioma, endometrial and ovarian
  carcinomas, squamous cell carcinoma of the head and neck, genitourinary
  carcinomas, and primary brain tumors}}.
\newblock {\emph{\JournalTitle{Adv. Anat. Pathol.}}}
  \textbf{\bibinfo{volume}{24}}, \bibinfo{pages}{311--335}
  (\bibinfo{year}{2017}).

\bibitem{kos2020pitfalls}
\bibinfo{author}{Kos, Z.} \emph{et~al.}
\newblock \bibinfo{journal}{\bibinfo{title}{Pitfalls in assessing stromal tumor
  infiltrating lymphocytes (stils) in breast cancer}}.
\newblock {\emph{\JournalTitle{NPJ breast cancer}}}
  \textbf{\bibinfo{volume}{6}}, \bibinfo{pages}{17} (\bibinfo{year}{2020}).

\bibitem{sato2023roles}
\bibinfo{author}{Sato, Y.}, \bibinfo{author}{Silina, K.},
  \bibinfo{author}{van~den Broek, M.}, \bibinfo{author}{Hirahara, K.} \&
  \bibinfo{author}{Yanagita, M.}
\newblock \bibinfo{journal}{\bibinfo{title}{The roles of tertiary lymphoid
  structures in chronic diseases}}.
\newblock {\emph{\JournalTitle{Nature Reviews Nephrology}}}
  \textbf{\bibinfo{volume}{19}}, \bibinfo{pages}{525--537}
  (\bibinfo{year}{2023}).

\bibitem{tilguidelines1}
\bibinfo{author}{Hendry, S.} \emph{et~al.}
\newblock \bibinfo{journal}{\bibinfo{title}{Assessing tumor-infiltrating
  lymphocytes in solid tumors: A practical review for pathologists and proposal
  for a standardized method from the international immunooncology biomarkers
  working group: Part 1: Assessing the host immune response, {TILs} in invasive
  breast carcinoma and ductal carcinoma in situ, metastatic tumor deposits and
  areas for further research}}.
\newblock {\emph{\JournalTitle{Adv. Anat. Pathol.}}}
  \textbf{\bibinfo{volume}{24}}, \bibinfo{pages}{235--251}
  (\bibinfo{year}{2017}).

\bibitem{niazi2019digital}
\bibinfo{author}{Niazi, M. K.~K.}, \bibinfo{author}{Parwani, A.~V.} \&
  \bibinfo{author}{Gurcan, M.~N.}
\newblock \bibinfo{journal}{\bibinfo{title}{Digital pathology and artificial
  intelligence}}.
\newblock {\emph{\JournalTitle{The lancet oncology}}}
  \textbf{\bibinfo{volume}{20}}, \bibinfo{pages}{e253--e261}
  (\bibinfo{year}{2019}).

\bibitem{backman2023spatial}
\bibinfo{author}{Backman, M.} \emph{et~al.}
\newblock \bibinfo{journal}{\bibinfo{title}{Spatial immunophenotyping of the
  tumour microenvironment in non--small cell lung cancer}}.
\newblock {\emph{\JournalTitle{European Journal of Cancer}}}
  \textbf{\bibinfo{volume}{185}}, \bibinfo{pages}{40--52}
  (\bibinfo{year}{2023}).

\bibitem{park2022artificial}
\bibinfo{author}{Park, S.} \emph{et~al.}
\newblock \bibinfo{journal}{\bibinfo{title}{Artificial intelligence--powered
  spatial analysis of tumor-infiltrating lymphocytes as complementary biomarker
  for immune checkpoint inhibition in non--small-cell lung cancer}}.
\newblock {\emph{\JournalTitle{Journal of Clinical Oncology}}}
  \textbf{\bibinfo{volume}{40}}, \bibinfo{pages}{1916--1928}
  (\bibinfo{year}{2022}).

\bibitem{spronck2022esmo}
\bibinfo{author}{Spronck, J.} \emph{et~al.}
\newblock \bibinfo{journal}{\bibinfo{title}{14p deep learning-based
  quantification of immune infiltrate for predicting response to pembrolizumab
  from pre-treatment biopsies of metastatic non-small cell lung cancer: A study
  on the pembro-rt phase ii trial}}.
\newblock {\emph{\JournalTitle{Immuno-Oncology and Technology}}}
  \textbf{\bibinfo{volume}{16}}, \bibinfo{pages}{100119},
  \url{https://doi.org/10.1016/j.iotech.2022.100119} (\bibinfo{year}{2022}).
\newblock \bibinfo{note}{Abstract Book of the ESMO Immuno-Oncology Congress
  2022 7-9 December 2022, Geneva Switzerland}.

\bibitem{kludt2024next}
\bibinfo{author}{Kludt, C.} \emph{et~al.}
\newblock \bibinfo{journal}{\bibinfo{title}{Next-generation lung cancer
  pathology: Development and validation of diagnostic and prognostic
  algorithms}}.
\newblock {\emph{\JournalTitle{Cell Reports Medicine}}}
  \textbf{\bibinfo{volume}{5}},
  \url{https://doi.org/10.1016/j.xcrm.2024.101697} (\bibinfo{year}{2024}).

\bibitem{van2024multi}
\bibinfo{author}{van Rijthoven, M.} \emph{et~al.}
\newblock \bibinfo{journal}{\bibinfo{title}{Multi-resolution deep learning
  characterizes tertiary lymphoid structures and their prognostic relevance in
  solid tumors}}.
\newblock {\emph{\JournalTitle{Communications Medicine}}}
  \textbf{\bibinfo{volume}{4}}, \bibinfo{pages}{5} (\bibinfo{year}{2024}).

\bibitem{li2020deep}
\bibinfo{author}{Li, Z.} \emph{et~al.}
\newblock \bibinfo{journal}{\bibinfo{title}{Deep learning methods for lung
  cancer segmentation in whole-slide histopathology images—the acdc@ lunghp
  challenge 2019}}.
\newblock {\emph{\JournalTitle{IEEE Journal of Biomedical and Health
  Informatics}}} \textbf{\bibinfo{volume}{25}}, \bibinfo{pages}{429--440},
  \url{https://doi.org/10.1109/JBHI.2020.3039741} (\bibinfo{year}{2020}).

\bibitem{verma2021monusac2020}
\bibinfo{author}{Verma, R.} \emph{et~al.}
\newblock \bibinfo{journal}{\bibinfo{title}{Monusac2020: A multi-organ nuclei
  segmentation and classification challenge}}.
\newblock {\emph{\JournalTitle{IEEE Transactions on Medical Imaging}}}
  \textbf{\bibinfo{volume}{40}}, \bibinfo{pages}{3413--3423},
  \url{https://doi.org/10.1109/TMI.2021.3085712} (\bibinfo{year}{2021}).

\bibitem{rkaczkowska2022deep}
\bibinfo{author}{R{\k{a}}czkowska, A.} \emph{et~al.}
\newblock \bibinfo{journal}{\bibinfo{title}{Deep learning-based tumor
  microenvironment segmentation is predictive of tumor mutations and patient
  survival in non-small-cell lung cancer}}.
\newblock {\emph{\JournalTitle{BMC cancer}}} \textbf{\bibinfo{volume}{22}},
  \bibinfo{pages}{1001} (\bibinfo{year}{2022}).

\bibitem{han2022wsss4luad}
\bibinfo{author}{Han, C.} \emph{et~al.}
\newblock \bibinfo{journal}{\bibinfo{title}{Wsss4luad: Grand challenge on
  weakly-supervised tissue semantic segmentation for lung adenocarcinoma}}.
\newblock {\emph{\JournalTitle{arXiv preprint arXiv:2204.06455}}}
  (\bibinfo{year}{2022}).

\bibitem{komura2023restaining}
\bibinfo{author}{Komura, D.} \emph{et~al.}
\newblock \bibinfo{journal}{\bibinfo{title}{Restaining-based annotation for
  cancer histology segmentation to overcome annotation-related limitations
  among pathologists}}.
\newblock {\emph{\JournalTitle{Patterns}}} \textbf{\bibinfo{volume}{4}}
  (\bibinfo{year}{2023}).

\bibitem{aubreville2023mitosis}
\bibinfo{author}{Aubreville, M.} \emph{et~al.}
\newblock \bibinfo{journal}{\bibinfo{title}{Mitosis domain generalization in
  histopathology images—the midog challenge}}.
\newblock {\emph{\JournalTitle{Medical Image Analysis}}}
  \textbf{\bibinfo{volume}{84}}, \bibinfo{pages}{102699}
  (\bibinfo{year}{2023}).

\bibitem{riihimaki2014metastatic}
\bibinfo{author}{Riihim{\"a}ki, M.} \emph{et~al.}
\newblock \bibinfo{journal}{\bibinfo{title}{Metastatic sites and survival in
  lung cancer}}.
\newblock {\emph{\JournalTitle{Lung cancer}}} \textbf{\bibinfo{volume}{86}},
  \bibinfo{pages}{78--84} (\bibinfo{year}{2014}).

\bibitem{tamura2015specific}
\bibinfo{author}{Tamura, T.} \emph{et~al.}
\newblock \bibinfo{journal}{\bibinfo{title}{Specific organ metastases and
  survival in metastatic non-small-cell lung cancer}}.
\newblock {\emph{\JournalTitle{Molecular and clinical oncology}}}
  \textbf{\bibinfo{volume}{3}}, \bibinfo{pages}{217--221}
  (\bibinfo{year}{2015}).

\bibitem{spronck2023nnunet}
\bibinfo{author}{Spronck, J.} \emph{et~al.}
\newblock \bibinfo{title}{nnunet meets pathology: bridging the gap for
  application to whole-slide images and computational biomarkers}.
\newblock In \emph{\bibinfo{booktitle}{Medical Imaging with Deep Learning}}
  (\bibinfo{year}{2023}).

\bibitem{van2024comparing}
\bibinfo{author}{van Eekelen, L.} \emph{et~al.}
\newblock \bibinfo{journal}{\bibinfo{title}{Comparing deep learning and
  pathologist quantification of cell-level pd-l1 expression in non-small cell
  lung cancer whole-slide images}}.
\newblock {\emph{\JournalTitle{Scientific Reports}}}
  \textbf{\bibinfo{volume}{14}}, \bibinfo{pages}{7136} (\bibinfo{year}{2024}).

\bibitem{Salgado2015til}
\bibinfo{author}{Salgado, R.} \emph{et~al.}
\newblock \bibinfo{journal}{\bibinfo{title}{The evaluation of
  tumor-infiltrating lymphocytes ({TILs}) in breast cancer: recommendations by
  an international {TILs} working group 2014}}.
\newblock {\emph{\JournalTitle{Ann. Oncol.}}} \textbf{\bibinfo{volume}{26}},
  \bibinfo{pages}{259--271} (\bibinfo{year}{2015}).

\bibitem{sedighzadeh2021macrophages}
\bibinfo{author}{Sedighzadeh, S.~S.}, \bibinfo{author}{Khoshbin, A.~P.},
  \bibinfo{author}{Razi, S.}, \bibinfo{author}{Keshavarz-Fathi, M.} \&
  \bibinfo{author}{Rezaei, N.}
\newblock \bibinfo{journal}{\bibinfo{title}{A narrative review of
  tumor-associated macrophages in lung cancer: regulation of macrophage
  polarization and therapeutic implications}}.
\newblock {\emph{\JournalTitle{Transl. Lung Cancer Res.}}}
  \textbf{\bibinfo{volume}{10}}, \bibinfo{pages}{1889--1916}
  (\bibinfo{year}{2021}).

\bibitem{litjens20181399}
\bibinfo{author}{Litjens, G.} \emph{et~al.}
\newblock \bibinfo{journal}{\bibinfo{title}{1399 h\&e-stained sentinel lymph
  node sections of breast cancer patients: the camelyon dataset}}.
\newblock {\emph{\JournalTitle{GigaScience}}} \textbf{\bibinfo{volume}{7}},
  \bibinfo{pages}{giy065} (\bibinfo{year}{2018}).

\bibitem{tigerpaper}
\bibinfo{author}{van Rijthoven, M.} \emph{et~al.}
\newblock \bibinfo{journal}{\bibinfo{title}{Tumor-infiltrating lymphocytes in
  breast cancer through artificial intelligence: biomarker analysis from the
  results of the tiger challenge}}.
\newblock {\emph{\JournalTitle{medRxiv}}} \bibinfo{pages}{2025--02}
  (\bibinfo{year}{2025}).

\bibitem{asap}
\bibinfo{author}{Litjens, G.}
\newblock \bibinfo{title}{Automate slide analysis platform (asap)}
  (\bibinfo{year}{2017}).

\bibitem{isensee2021nnu}
\bibinfo{author}{Isensee, F.}, \bibinfo{author}{Jaeger, P.~F.},
  \bibinfo{author}{Kohl, S.~A.}, \bibinfo{author}{Petersen, J.} \&
  \bibinfo{author}{Maier-Hein, K.~H.}
\newblock \bibinfo{journal}{\bibinfo{title}{nnu-net: a self-configuring method
  for deep learning-based biomedical image segmentation}}.
\newblock {\emph{\JournalTitle{Nature methods}}} \textbf{\bibinfo{volume}{18}},
  \bibinfo{pages}{203--211} (\bibinfo{year}{2021}).

\bibitem{glenn_jocher_2022_7347926}
\bibinfo{author}{Jocher, G.} \emph{et~al.}
\newblock \bibinfo{title}{{ultralytics/yolov5: v7.0 - YOLOv5 SOTA Realtime
  Instance Segmentation}}, \url{10.5281/zenodo.7347926} (\bibinfo{year}{2022}).

\bibitem{ignitetoolkit}
\bibinfo{author}{Spronck, J.} \emph{et~al.}
\newblock \bibinfo{title}{Ignite data toolkit: a tissue and cell-level
  annotated h\&e and pd-l1 histopathology image dataset in non-small cell lung
  cancer}, \url{10.5281/zenodo.15674785} (\bibinfo{year}{2025}).

\bibitem{lin2014microsoft}
\bibinfo{author}{Lin, T.-Y.} \emph{et~al.}
\newblock \bibinfo{title}{Microsoft coco: Common objects in context}.
\newblock In \emph{\bibinfo{booktitle}{Computer Vision--ECCV 2014: 13th
  European Conference, Zurich, Switzerland, September 6-12, 2014, Proceedings,
  Part V 13}}, \bibinfo{pages}{740--755} (\bibinfo{organization}{Springer},
  \bibinfo{year}{2014}).

\end{thebibliography}

\end{document}